
%
\expandafter\ifx\csname phyzzx\endcsname\relax\else
 \errhelp{Hit <CR> and go ahead.}
 \errmessage{PHYZZX macros are already loaded or input. }
 \endinput \fi
\catcode`\@=11 
%
%
%
\font\seventeenrm=cmr17
\font\fourteenrm=cmr12 scaled\magstep1
\font\twelverm=cmr12
\font\ninerm=cmr9            \font\sixrm=cmr6
%
\font\fourteenbf=cmbx10 scaled\magstep2
\font\twelvebf=cmbx12
\font\ninebf=cmbx9            \font\sixbf=cmbx6
%
\font\fourteeni=cmmi10 scaled\magstep2      \skewchar\fourteeni='177
\font\twelvei=cmmi12			        \skewchar\twelvei='177
\font\ninei=cmmi9                           \skewchar\ninei='177
\font\sixi=cmmi6                            \skewchar\sixi='177
%
\font\fourteensy=cmsy10 scaled\magstep2     \skewchar\fourteensy='60
\font\twelvesy=cmsy10 scaled\magstep1	    \skewchar\twelvesy='60
\font\ninesy=cmsy9                          \skewchar\ninesy='60
\font\sixsy=cmsy6                           \skewchar\sixsy='60
%
\font\fourteenex=cmex10 scaled\magstep2
\font\twelveex=cmex10 scaled\magstep1
%
\font\fourteensl=cmsl12 scaled\magstep1
\font\twelvesl=cmsl12
\font\ninesl=cmsl9
%
\font\fourteenit=cmti12 scaled\magstep1
\font\twelveit=cmti12
\font\nineit=cmti9
\font\fourteentt=cmtt10 scaled\magstep2
\font\twelvett=cmtt12
\font\fourteencp=cmcsc10 scaled\magstep2
\font\twelvecp=cmcsc10 scaled\magstep1
\font\tencp=cmcsc10
\newfam\cpfam
\newdimen\b@gheight		\b@gheight=12pt
\newcount\f@ntkey		\f@ntkey=0
\def\f@m{\afterassignment\samef@nt\f@ntkey=}
\def\samef@nt{\fam=\f@ntkey \the\textfont\f@ntkey\relax}
\def\rm{\f@m0 }
\def\mit{\f@m1 }         
\def\cal{\f@m2 }
\def\it{\f@m\itfam}
\def\sl{\f@m\slfam}
\def\bf{\f@m\bffam}
\def\tt{\f@m\ttfam}
\def\caps{\f@m\cpfam}
\def\fourteenpoint{\relax
    \textfont0=\fourteenrm          \scriptfont0=\tenrm
      \scriptscriptfont0=\sevenrm
    \textfont1=\fourteeni           \scriptfont1=\teni
      \scriptscriptfont1=\seveni
    \textfont2=\fourteensy          \scriptfont2=\tensy
      \scriptscriptfont2=\sevensy
    \textfont3=\fourteenex          \scriptfont3=\twelveex
      \scriptscriptfont3=\tenex
    \textfont\itfam=\fourteenit     \scriptfont\itfam=\tenit
    \textfont\slfam=\fourteensl     \scriptfont\slfam=\tensl
    \textfont\bffam=\fourteenbf     \scriptfont\bffam=\tenbf
      \scriptscriptfont\bffam=\sevenbf
    \textfont\ttfam=\fourteentt
    \textfont\cpfam=\fourteencp
    \samef@nt
    \b@gheight=14pt
    \setbox\strutbox=\hbox{\vrule height 0.85\b@gheight
				depth 0.35\b@gheight width\z@ }}
\def\twelvepoint{\relax
    \textfont0=\twelverm          \scriptfont0=\ninerm
      \scriptscriptfont0=\sixrm
    \textfont1=\twelvei           \scriptfont1=\ninei
      \scriptscriptfont1=\sixi
    \textfont2=\twelvesy           \scriptfont2=\ninesy
      \scriptscriptfont2=\sixsy
    \textfont3=\twelveex          \scriptfont3=\tenex
      \scriptscriptfont3=\tenex
    \textfont\itfam=\twelveit     \scriptfont\itfam=\nineit
    \textfont\slfam=\twelvesl     \scriptfont\slfam=\ninesl
    \textfont\bffam=\twelvebf     \scriptfont\bffam=\ninebf
      \scriptscriptfont\bffam=\sixbf
    \textfont\ttfam=\twelvett
    \textfont\cpfam=\twelvecp
    \samef@nt
    \b@gheight=12pt
    \setbox\strutbox=\hbox{\vrule height 0.85\b@gheight
				depth 0.35\b@gheight width\z@ }}
\def\tenpoint{\relax
    \textfont0=\tenrm          \scriptfont0=\sevenrm
      \scriptscriptfont0=\fiverm
    \textfont1=\teni           \scriptfont1=\seveni
      \scriptscriptfont1=\fivei
    \textfont2=\tensy          \scriptfont2=\sevensy
      \scriptscriptfont2=\fivesy
    \textfont3=\tenex          \scriptfont3=\tenex
      \scriptscriptfont3=\tenex
    \textfont\itfam=\tenit     \scriptfont\itfam=\seveni
    \textfont\slfam=\tensl     \scriptfont\slfam=\sevenrm
    \textfont\bffam=\tenbf     \scriptfont\bffam=\sevenbf
      \scriptscriptfont\bffam=\fivebf
    \textfont\ttfam=\tentt
    \textfont\cpfam=\tencp
    \samef@nt
    \b@gheight=10pt
    \setbox\strutbox=\hbox{\vrule height 0.85\b@gheight
				depth 0.35\b@gheight width\z@ }}
%
%
%
\normalbaselineskip = 20pt plus 0.2pt minus 0.1pt
\normallineskip = 1.5pt plus 0.1pt minus 0.1pt
\normallineskiplimit = 1.5pt
\newskip\normaldisplayskip
\normaldisplayskip = 20pt plus 5pt minus 10pt
\newskip\normaldispshortskip
\normaldispshortskip = 6pt plus 5pt
\newskip\normalparskip
\normalparskip = 6pt plus 2pt minus 1pt
\newskip\skipregister
\skipregister = 5pt plus 2pt minus 1.5pt
\newif\ifsingl@    \newif\ifdoubl@
\newif\iftwelv@    \twelv@true
\def\singlespace{\singl@true\doubl@false\spaces@t}
\def\doublespace{\singl@false\doubl@true\spaces@t}
\def\normalspace{\singl@false\doubl@false\spaces@t}
\def\Tenpoint{\tenpoint\twelv@false\spaces@t}
\def\Twelvepoint{\twelvepoint\twelv@true\spaces@t}
\def\spaces@t{\relax
      \iftwelv@ \ifsingl@\subspaces@t3:4;\else\subspaces@t1:1;\fi
       \else \ifsingl@\subspaces@t3:5;\else\subspaces@t4:5;\fi \fi
      \ifdoubl@ \multiply\baselineskip by 5
         \divide\baselineskip by 4 \fi }
\def\subspaces@t#1:#2;{
      \baselineskip = \normalbaselineskip
      \multiply\baselineskip by #1 \divide\baselineskip by #2
      \lineskip = \normallineskip
      \multiply\lineskip by #1 \divide\lineskip by #2
      \lineskiplimit = \normallineskiplimit
      \multiply\lineskiplimit by #1 \divide\lineskiplimit by #2
      \parskip = \normalparskip
      \multiply\parskip by #1 \divide\parskip by #2
      \abovedisplayskip = \normaldisplayskip
      \multiply\abovedisplayskip by #1 \divide\abovedisplayskip by #2
      \belowdisplayskip = \abovedisplayskip
      \abovedisplayshortskip = \normaldispshortskip
      \multiply\abovedisplayshortskip by #1
        \divide\abovedisplayshortskip by #2
      \belowdisplayshortskip = \abovedisplayshortskip
      \advance\belowdisplayshortskip by \belowdisplayskip
      \divide\belowdisplayshortskip by 2
      \smallskipamount = \skipregister
      \multiply\smallskipamount by #1 \divide\smallskipamount by #2
      \medskipamount = \smallskipamount \multiply\medskipamount by 2
      \bigskipamount = \smallskipamount \multiply\bigskipamount by 4 }
\def\normalbaselines{ \baselineskip=\normalbaselineskip
   \lineskip=\normallineskip \lineskiplimit=\normallineskip
   \iftwelv@\else \multiply\baselineskip by 4 \divide\baselineskip by 5
     \multiply\lineskiplimit by 4 \divide\lineskiplimit by 5
     \multiply\lineskip by 4 \divide\lineskip by 5 \fi }
\Twelvepoint  
\interlinepenalty=50
\interfootnotelinepenalty=5000
\predisplaypenalty=9000
\postdisplaypenalty=500
\hfuzz=1pt
\vfuzz=0.2pt
\voffset=0pt
\dimen\footins=8 truein
%
%
%
\def\pagecontents{
   \ifvoid\topins\else\unvbox\topins\vskip\skip\topins\fi
   \dimen@ = \dp255 \unvbox255
   \ifvoid\footins\else\vskip\skip\footins\footrule\unvbox\footins\fi
   \ifr@ggedbottom \kern-\dimen@ \vfil \fi }
\def\makeheadline{\vbox to 0pt{ \skip@=\topskip
      \advance\skip@ by -12pt \advance\skip@ by -2\normalbaselineskip
      \vskip\skip@ \line{\vbox to 12pt{}\the\headline} \vss
      }\nointerlineskip}
\def\makefootline{\baselineskip = 1.5\normalbaselineskip
                 \line{\the\footline}}
\newif\iffrontpage
\newif\ifletterstyle
\newif\ifp@genum
\def\nopagenumbers{\p@genumfalse}
\def\pagenumbers{\p@genumtrue}
\pagenumbers
\newtoks\paperheadline
\newtoks\letterheadline
\newtoks\paperfootline
\newtoks\letterfootline
\newtoks\letterinfo
\newtoks\Letterinfo
\newtoks\date
\footline={\ifletterstyle\the\letterfootline\else\the\paperfootline\fi}
\paperfootline={\hss\iffrontpage\else\ifp@genum\tenrm\folio\hss\fi\fi}
\letterfootline={\iffrontpage\LETTERFOOT\else\hfil\fi}
\Letterinfo={\hfil}
\letterinfo={\hfil}
\def\LETTERFOOT{\hfil} 
%
\def\LETTERHEAD{\vtop{\baselineskip=20pt\hbox to
\hsize{\hfil\seventeenrm\strut 
CALIFORNIA INSTITUTE OF TECHNOLOGY \hfil}
\hbox to \hsize{\hfil\ninerm\strut
CHARLES C. LAURITSEN LABORATORY OF HIGH ENERGY PHYSICS \hfil}
\hbox to \hsize{\hfil\ninerm\strut
PASADENA, CALIFORNIA 91125 \hfil}}}
\headline={\ifletterstyle\the\letterheadline\else\the\paperheadline\fi}
\paperheadline={\hfil}
\letterheadline{\iffrontpage \LETTERHEAD\else
    \rm \ifp@genum \hfil \folio\hfil\fi\fi}
\def\monthname{\relax\ifcase\month 0/\or January\or February\or
   March\or April\or May\or June\or July\or August\or September\or
   October\or November\or December\else\number\month/\fi}
\def\today{\monthname\ \number\day, \number\year}
\date={\today}
\countdef\pageno=1      \countdef\pagen@=0
\countdef\pagenumber=1  \pagenumber=1
\def\advancepageno{\global\advance\pagen@ by 1
   \ifnum\pagenumber<0 \global\advance\pagenumber by -1
    \else\global\advance\pagenumber by 1 \fi \global\frontpagefalse }
\def\folio{\ifnum\pagenumber<0 \romannumeral-\pagenumber
           \else \number\pagenumber \fi }
\def\footrule{\dimen@=\prevdepth\nointerlineskip
   \vbox to 0pt{\vskip -0.25\baselineskip \hrule width 0.35\hsize \vss}
   \prevdepth=\dimen@ }
\newtoks\foottokens
\foottokens={}
\newdimen\footindent
\footindent=24pt
\def\vfootnote#1{\insert\footins\bgroup  
   \interlinepenalty=\interfootnotelinepenalty \floatingpenalty=20000
   \singl@true\doubl@false\Tenpoint
   \splittopskip=\ht\strutbox \boxmaxdepth=\dp\strutbox
   \leftskip=\footindent \rightskip=\z@skip
   \parindent=0.5\footindent \parfillskip=0pt plus 1fil
   \spaceskip=\z@skip \xspaceskip=\z@skip
   \the\foottokens
   \Textindent{$ #1 $}\footstrut\futurelet\next\fo@t}
\def\Textindent#1{\noindent\llap{#1\enspace}\ignorespaces}
\def\footnote#1{\attach{#1}\vfootnote{#1}}

\let\footsymbol=\star
\newcount\lastf@@t           \lastf@@t=-1
\newcount\footsymbolcount    \footsymbolcount=0
\newif\ifPhysRev
\def\bumpfootsymbolcount{\relax
   \iffrontpage \bumpfootsymbolNP \else \advance\lastf@@t by 1
     \ifPhysRev \bumpfootsymbolPR \else \bumpfootsymbolNP \fi \fi
   \global\lastf@@t=\pagen@ }
\def\bumpfootsymbolNP{\ifnum\footsymbolcount <0 \global\footsymbolcount =0 \fi
    \ifnum\lastf@@t<\pagen@ \global\footsymbolcount=0
     \else \global\advance\footsymbolcount by 1 \fi }
\def\bumpfootsymbolPR{\ifnum\footsymbolcount >0 \global\footsymbolcount =0 \fi
      \global\advance\footsymbolcount by -1 }
\def\fd@f#1 {\xdef\footsymbol{\mathchar"#1 }}
\def\generatefootsymbol{\ifcase\footsymbolcount \fd@f 13F \or \fd@f 279
	\or \fd@f 27A \or \fd@f 278 \or \fd@f 27B \else
	\ifnum\footsymbolcount <0 \fd@f{023 \number-\footsymbolcount }
	 \else \fd@f 203 {\loop \ifnum\footsymbolcount >5
		\fd@f{203 \footsymbol } \advance\footsymbolcount by -1
		\repeat }\fi \fi }

\def\nonfrenchspacing{\sfcode`\.=3001 \sfcode`\!=3000 \sfcode`\?=3000
	\sfcode`\:=2000 \sfcode`\;=1500 \sfcode`\,=1251 }
\nonfrenchspacing
\newdimen\d@twidth
{\setbox0=\hbox{s.} \global\d@twidth=\wd0 \setbox0=\hbox{s}
	\global\advance\d@twidth by -\wd0 }
\def\removehglue{\loop \unskip \ifdim\lastskip >\z@ \repeat }
\def\roll@ver#1{\removehglue \nobreak \count255 =\spacefactor \dimen@=\z@
	\ifnum\count255 =3001 \dimen@=\d@twidth \fi
	\ifnum\count255 =1251 \dimen@=\d@twidth \fi
    \iftwelv@ \kern-\dimen@ \else \kern-0.83\dimen@ \fi
   #1\spacefactor=\count255 }
\def\step@ver#1{\relax \ifmmode #1\else \ifhmode
	\roll@ver{${}#1$}\else {\setbox0=\hbox{${}#1$}}\fi\fi }
\def\attach#1{\step@ver{\strut^{\mkern 2mu #1} }}
%
%
%
\newcount\chapternumber      \chapternumber=0
\newcount\sectionnumber      \sectionnumber=0
\newcount\equanumber         \equanumber=0
\let\chapterlabel=\relax
\let\sectionlabel=\relax
\newtoks\chapterstyle        \chapterstyle={\Number}
\newtoks\sectionstyle        \sectionstyle={\chapterlabel\Number}
\newskip\chapterskip         \chapterskip=\bigskipamount
\newskip\sectionskip         \sectionskip=\medskipamount
\newskip\headskip            \headskip=8pt plus 3pt minus 3pt
\newdimen\chapterminspace    \chapterminspace=15pc
\newdimen\sectionminspace    \sectionminspace=10pc
\newdimen\referenceminspace  \referenceminspace=25pc
\def\chapterreset{\global\advance\chapternumber by 1
   \ifnum\equanumber<0 \else\global\equanumber=0\fi
   \sectionnumber=0 \makechapterlabel}
\def\makechapterlabel{\let\sectionlabel=\relax
   \xdef\chapterlabel{\the\chapterstyle{\the\chapternumber}.}}
\def\alphabetic#1{\count255='140 \advance\count255 by #1\char\count255}
\def\Alphabetic#1{\count255='100 \advance\count255 by #1\char\count255}
\def\Roman#1{\uppercase\expandafter{\romannumeral #1}}
\def\roman#1{\romannumeral #1}
\def\Number#1{\number #1}
\def\BLANC#1{}
\def\titlestyle#1{\par\begingroup \interlinepenalty=9999
     \leftskip=0.02\hsize plus 0.23\hsize minus 0.02\hsize
     \rightskip=\leftskip \parfillskip=0pt
     \hyphenpenalty=9000 \exhyphenpenalty=9000
     \tolerance=9999 \pretolerance=9000
     \spaceskip=0.333em \xspaceskip=0.5em
     \iftwelv@\fourteenpoint\else\twelvepoint\fi
   \noindent #1\par\endgroup }
\def\spacecheck#1{\dimen@=\pagegoal\advance\dimen@ by -\pagetotal
   \ifdim\dimen@<#1 \ifdim\dimen@>0pt \vfil\break \fi\fi}
\def\TableOfContentEntry#1#2#3{\relax}
\def\chapter#1{\par \penalty-300 \vskip\chapterskip
   \spacecheck\chapterminspace
   \chapterreset \titlestyle{\chapterlabel\ #1}
   \TableOfContentEntry c\chapterlabel{#1}
   \nobreak\vskip\headskip \penalty 30000
   \wlog{\string\chapter\space \chapterlabel} }

\def\section#1{\par \ifnum\the\lastpenalty=30000\else
   \penalty-200\vskip\sectionskip \spacecheck\sectionminspace\fi
   \global\advance\sectionnumber by 1
   \xdef\sectionlabel{\the\sectionstyle\the\sectionnumber}
   \wlog{\string\section\space \sectionlabel}
   \TableOfContentEntry s\sectionlabel{#1}
   \noindent {\caps\enspace\sectionlabel\quad #1}\par
   \nobreak\vskip\headskip \penalty 30000 }
\def\subsection#1{\par
   \ifnum\the\lastpenalty=30000\else \penalty-100\smallskip \fi
   \noindent\undertext{#1}\enspace \vadjust{\penalty5000}}

\def\undertext#1{\vtop{\hbox{#1}\kern 1pt \hrule}}
\def\ack{\par\penalty-100\medskip \spacecheck\sectionminspace
   \line{\fourteenrm\hfil ACKNOWLEDGEMENTS\hfil}\nobreak\vskip\headskip }
\def\APPENDIX#1#2{\par\penalty-300\vskip\chapterskip
   \spacecheck\chapterminspace \chapterreset \xdef\chapterlabel{#1}
   \titlestyle{APPENDIX #2} \nobreak\vskip\headskip \penalty 30000
   \TableOfContentEntry a{#1}{#2}
   \wlog{\string\Appendix\ \chapterlabel} }
\def\Appendix#1{\APPENDIX{#1}{#1}}
\def\appendix{\APPENDIX{A}{}}
\def\unnumberedchapters{\let\makechapterlabel=\relax \let\chapterlabel=\relax
   \sectionstyle={\BLANC}\let\sectionlabel=\relax \sequentialequations }
%
%
%
\def\eqname#1{\relax \ifnum\equanumber<0
     \xdef#1{{\noexpand\rm(\number-\equanumber)}}%
       \global\advance\equanumber by -1
    \else \global\advance\equanumber by 1
      \xdef#1{{\noexpand\rm(\chapterlabel\number\equanumber)}} \fi #1}

\def\eqn{\eqno\eqname}

\def\eqinsert#1{\noalign{\dimen@=\prevdepth \nointerlineskip
   \setbox0=\hbox to\displaywidth{\hfil #1}
   \vbox to 0pt{\kern 0.5\baselineskip\hbox{$\!\box0\!$}\vss}
   \prevdepth=\dimen@}}
%

%
%
\def\GENITEM#1;#2{\par \hangafter=0 \hangindent=#1
    \Textindent{$ #2 $}\ignorespaces}
\outer\def\newitem#1=#2;{\gdef#1{\GENITEM #2;}}
\newdimen\itemsize                \itemsize=30pt
\newitem\item=1\itemsize;
\newitem\sitem=1.75\itemsize;     
\newitem\ssitem=2.5\itemsize;     
\outer\def\newlist#1=#2&#3&#4;{\toks0={#2}\toks1={#3}%
   \count255=\escapechar \escapechar=-1
   \alloc@0\list\countdef\insc@unt\listcount     \listcount=0
   \edef#1{\par
      \countdef\listcount=\the\allocationnumber
      \advance\listcount by 1
      \hangafter=0 \hangindent=#4
      \Textindent{\the\toks0{\listcount}\the\toks1}}
   \expandafter\expandafter\expandafter
    \edef\c@t#1{begin}{\par
      \countdef\listcount=\the\allocationnumber \listcount=1
      \hangafter=0 \hangindent=#4
      \Textindent{\the\toks0{\listcount}\the\toks1}}
   \expandafter\expandafter\expandafter
    \edef\c@t#1{con}{\par \hangafter=0 \hangindent=#4 \noindent}
   \escapechar=\count255}
\def\c@t#1#2{\csname\string#1#2\endcsname}
\newlist\point=\Number&.&1.0\itemsize;
\newlist\subpoint=(\alphabetic&)&1.75\itemsize;
\newlist\subsubpoint=(\roman&)&2.5\itemsize;
%

%
%
%
%
\newcount\referencecount     \referencecount=0
\newcount\lastrefsbegincount \lastrefsbegincount=0
\newif\ifreferenceopen       \newwrite\referencewrite
\newif\ifrw@trailer
\newdimen\refindent     \refindent=30pt
\def\NPrefmark#1{\attach{\scriptscriptstyle [ #1 ] }}
\let\PRrefmark=\attach
\def\refmark#1{\relax\ifPhysRev\PRrefmark{#1}\else\NPrefmark{#1}\fi}
\def\refend@{\refmark{\number\referencecount}}
\def\refend{\refend@{}\space }
\def\refsend{\refmark{\count255=\referencecount
   \advance\count255 by-\lastrefsbegincount
   \ifcase\count255 \number\referencecount
   \or \number\lastrefsbegincount,\number\referencecount
   \else \number\lastrefsbegincount-\number\referencecount \fi}\space }
\def\refitem#1{\par \hangafter=0 \hangindent=\refindent \Textindent{#1}}
\def\Ref{\rw@trailertrue\REF}
\def\ref{\Ref\?}

\def\REF#1{\r@fstart{#1}%
   \rw@begin{\the\referencecount.}\rw@end}
\def\REFS#1{\r@fstart{#1}%
   \lastrefsbegincount=\referencecount
   \rw@begin{\the\referencecount.}\rw@end}
\def\r@fstart#1{\chardef\rw@write=\referencewrite \let\rw@ending=\refend@
   \ifreferenceopen \else \global\referenceopentrue
   \immediate\openout\referencewrite=referenc.txa
   \toks0={\catcode`\^^M=10}\immediate\write\rw@write{\the\toks0} \fi
   \global\advance\referencecount by 1 \xdef#1{\the\referencecount}}
{\catcode`\^^M=\active %
 \gdef\rw@begin#1{\immediate\write\rw@write{\noexpand\refitem{#1}}%
   \begingroup \catcode`\^^M=\active \let^^M=\relax}%
 \gdef\rw@end#1{\rw@@end #1^^M\rw@terminate \endgroup%
   \ifrw@trailer\rw@ending\global\rw@trailerfalse\fi }%
 \gdef\rw@@end#1^^M{\toks0={#1}\immediate\write\rw@write{\the\toks0}%
   \futurelet\n@xt\rw@test}%
 \gdef\rw@test{\ifx\n@xt\rw@terminate \let\n@xt=\relax%
       \else \let\n@xt=\rw@@end \fi \n@xt}%
}
\let\rw@ending=\relax
\let\rw@terminate=\relax
\let\splitout=\relax
\def\Closeout#1{\toks0={\catcode`\^^M=5}\immediate\write#1{\the\toks0}%
   \immediate\closeout#1}
%
%
\newcount\figurecount     \figurecount=0
\newcount\tablecount      \tablecount=0
\newif\iffigureopen       \newwrite\figurewrite
\newif\iftableopen        \newwrite\tablewrite
\def\FIG#1{\f@gstart{#1}%
   \rw@begin{\the\figurecount)}\rw@end}

\def\Fig{\rw@trailertrue\def\rw@ending{Fig.~\?}\FIG\?}
\def\fig{\rw@trailertrue\def\rw@ending{fig.~\?}\FIG\?}
\def\TABLE#1{\T@Bstart{#1}%
   \rw@begin{\the\tableecount:}\rw@end}
\def\Table{\rw@trailertrue\def\rw@ending{Table~\?}\TABLE\?}
\def\f@gstart#1{\chardef\rw@write=\figurewrite
   \iffigureopen \else \global\figureopentrue
   \immediate\openout\figurewrite=figures.txa
   \toks0={\catcode`\^^M=10}\immediate\write\rw@write{\the\toks0} \fi
   \global\advance\figurecount by 1 \xdef#1{\the\figurecount}}
\def\T@Bstart#1{\chardef\rw@write=\tablewrite
   \iftableopen \else \global\tableopentrue
   \immediate\openout\tablewrite=tables.txa
   \toks0={\catcode`\^^M=10}\immediate\write\rw@write{\the\toks0} \fi
   \global\advance\tablecount by 1 \xdef#1{\the\tablecount}}
%

%
%
%
\def\getfigure#1{\global\advance\figurecount by 1
   \xdef#1{\the\figurecount}\count255=\escapechar \escapechar=-1
   \edef\n@xt{\noexpand\g@tfigure\csname\string#1Body\endcsname}%
   \escapechar=\count255 \n@xt }
\def\g@tfigure#1#2 {\errhelp=\disabledfigures \let#1=\relax
   \errmessage{\string\getfigure\space disabled}}
\newhelp\disabledfigures{ Empty figure of zero size assumed.}
\def\figinsert#1{\midinsert\Tenpoint\medskip
   \count255=\escapechar \escapechar=-1
   \edef\n@xt{\csname\string#1Body\endcsname}
   \escapechar=\count255 \centerline{\n@xt}
   \bigskip\narrower\narrower
   \noindent{\it Figure}~#1.\quad }
%
%
%
\def\masterreset{\global\pagenumber=1 \global\chapternumber=0
   \global\equanumber=0 \global\sectionnumber=0
   \global\referencecount=0 \global\figurecount=0 \global\tablecount=0 }
\def\FRONTPAGE{\ifvoid255\else\vfill\penalty-20000\fi
      \masterreset\global\frontpagetrue
      \global\lastf@@t=0 \global\footsymbolcount=0}

\def\paperstyle{\letterstylefalse\normalspace\papersize}
\def\letterstyle{\letterstyletrue\singlespace\lettersize}
\def\papersize{\hsize=35 truepc\vsize=50 truepc\hoffset=-2.51688 truepc
               \skip\footins=\bigskipamount}
\def\lettersize{\hsize=5.5 truein\vsize=8.25 truein\hoffset=.4875 truein
	\voffset=.3125 truein
   \skip\footins=\smallskipamount \multiply\skip\footins by 3 }
\paperstyle   
%
%
\def\MEMO{\letterstyle \letterinfo={\hfil } \let\rule=\memorule
	\FRONTPAGE \memohead }
\let\memohead=\relax

\def\memit@m#1{\smallskip \hangafter=0 \hangindent=1in
      \Textindent{\caps #1}}
\def\subject{\memit@m{Subject:}}
\def\topic{\memit@m{Topic:}}
\def\from{\memit@m{From:}}
\def\to{\relax \ifmmode \rightarrow \else \memit@m{To:}\fi }
\def\memorule{\medskip\hrule height 1pt\bigskip}
\newwrite\labelswrite
\newtoks\rw@toks

\def\addressee#1{\null\vskip .5truein\line{
\hskip 0.5\hsize minus 0.5\hsize\the\date\hfil}\bigskip
   \ialign to\hsize{\strut ##\hfil\tabskip 0pt plus \hsize \cr #1\crcr}
   \writelabel{#1}\medskip\par\noindent}
\def\rwl@begin#1\cr{\rw@toks={#1\crcr}\relax
   \immediate\write\labelswrite{\the\rw@toks}\futurelet\n@xt\rwl@next}
\def\rwl@next{\ifx\n@xt\rwl@end \let\n@xt=\relax
      \else \let\n@xt=\rwl@begin \fi \n@xt}
\let\rwl@end=\relax
\def\writelabel#1{\immediate\write\labelswrite{\noexpand\labelbegin}
     \rwl@begin #1\cr\rwl@end
     \immediate\write\labelswrite{\noexpand\labelend}}
\newbox\FromLabelBox

\let\labelend=\relax
\def\labelbegin#1\labelend{\setbox0=\vbox{\ialign{##\hfil\cr #1\crcr}}
     \MakeALabel }
\newtoks\FromAddress
\FromAddress={}
\def\MakeFromBox#1{\global\setbox\FromLabelBox=\vbox{\Tenpoint
     \ialign{##\hfil\cr #1\the\FromAddress\crcr}}}
\newdimen\labelwidth		\labelwidth=6in
\def\MakeALabel{\vskip 1pt \hbox{\vrule \vbox{
	\hsize=\labelwidth \hrule\bigskip
	\leftline{\hskip 1\parindent \copy\FromLabelBox}\bigskip
	\centerline{\hfil \box0 } \bigskip \hrule
	}\vrule } \vskip 1pt plus 1fil }
\newskip\signatureskip       \signatureskip=30pt
\def\signed#1{\par \penalty 9000 \medskip \dt@pfalse
  \everycr={\noalign{\ifdt@p\vskip\signatureskip\global\dt@pfalse\fi}}
  \setbox0=\vbox{\singlespace \ialign{\strut ##\hfil\crcr
   \noalign{\global\dt@ptrue}#1\crcr}}
  \line{\hskip 0.5\hsize minus 0.5\hsize \box0\hfil} \medskip }
\newbox\letterb@x
\def\lettertext{\par\unvcopy\letterb@x\par}
\def\multiletter{\setbox\letterb@x=\vbox\bgroup
      \everypar{\vrule height 1\baselineskip depth 0pt width 0pt }
      \singlespace \topskip=\baselineskip }
\def\letterend{\par\egroup}
%
%
%
\newskip\frontpageskip
\newtoks\Pubnum
\newtoks\pubtype
\newif\ifp@bblock  \p@bblocktrue
\def\PH@SR@V{\doubl@true \baselineskip=24.1pt plus 0.2pt minus 0.1pt
             \parskip= 3pt plus 2pt minus 1pt }
\def\PHYSREV{\paperstyle\PhysRevtrue\PH@SR@V}
\def\titlepage{\FRONTPAGE\paperstyle\ifPhysRev\PH@SR@V\fi
   \ifp@bblock\p@bblock \else\hrule height\z@ \relax \fi }
\def\nopubblock{\p@bblockfalse}

\frontpageskip=12pt plus .5fil minus 2pt
\pubtype={\tensl Preliminary Version}
\Pubnum={}
\def\p@bblock{\begingroup \tabskip=\hsize minus \hsize
   \baselineskip=1.5\ht\strutbox \topspace-2\baselineskip
   \halign to\hsize{\strut ##\hfil\tabskip=0pt\crcr
       \the\Pubnum\crcr\the\date\crcr\the\pubtype\crcr}\endgroup}
\def\title#1{\vskip\frontpageskip \titlestyle{#1} \vskip\headskip }
\def\author#1{\vskip\frontpageskip\titlestyle{\twelvecp #1}\nobreak}

\def\address#1{\par\kern 5pt\titlestyle{\twelvepoint\it #1}}
\def\andaddress{\par\kern 5pt \centerline{\sl and} \address}

\def\abstract{\par\dimen@=\prevdepth \hrule height\z@ \prevdepth=\dimen@
   \vskip\frontpageskip\centerline{\fourteenrm ABSTRACT}\vskip\headskip }

%
%
%

\def\\{\relax \ifmmode \backslash \else {\tt\char`\\}\fi }
\def\sequentialequations{\relax\if\equanumber<0\else\global\equanumber=-1\fi}

\def\journal#1&#2(#3){\unskip, \sl #1\unskip~\bf\ignorespaces #2\rm (19#3),}

\def\topspace{\hrule height 0pt depth 0pt \vskip}

\def\Buildrel#1\under#2{\mathrel{\mathop{#2}\limits_{#1}}}
\def\becomes#1{\mathchoice{\becomes@\scriptstyle{#1}}{\becomes@\scriptstyle
   {#1}}{\becomes@\scriptscriptstyle{#1}}{\becomes@\scriptscriptstyle{#1}}}
\def\becomes@#1#2{\mathrel{\setbox0=\hbox{$\m@th #1{\,#2\,}$}%
	\mathop{\hbox to \wd0 {\rightarrowfill}}\limits_{#2}}}

\def\ket#1{\left| #1\right\rangle}
\def\VEV#1{\left\langle #1\right\rangle}

\let\int=\intop         
\def\lsim{\mathrel{\mathpalette\@versim<}}
\def\gsim{\mathrel{\mathpalette\@versim>}}
\def\@versim#1#2{\vcenter{\offinterlineskip
	\ialign{$\m@th#1\hfil##\hfil$\crcr#2\crcr\sim\crcr } }}
\def\big#1{{\hbox{$\left#1\vbox to 0.85\b@gheight{}\right.\n@space$}}}
\def\Big#1{{\hbox{$\left#1\vbox to 1.15\b@gheight{}\right.\n@space$}}}
\def\bigg#1{{\hbox{$\left#1\vbox to 1.45\b@gheight{}\right.\n@space$}}}
\def\Bigg#1{{\hbox{$\left#1\vbox to 1.75\b@gheight{}\right.\n@space$}}}
%
%
%
\let\sec@nt=\sec
\def\sec{\relax\ifmmode\let\n@xt=\sec@nt\else\let\n@xt\section\fi\n@xt}
\def\obsolete#1{\message{Macro \string #1 is obsolete.}}
\def\firstsec#1{\obsolete\firstsec \section{#1}}
\def\firstsubsec#1{\obsolete\firstsubsec \subsection{#1}}
\def\thispage#1{\obsolete\thispage \global\pagenumber=#1\frontpagefalse}
\def\thischapter#1{\obsolete\thischapter \global\chapternumber=#1}
\def\REFSCON{\obsolete\REFSCON\REF}
\def\splitout{\obsolete\splitout\relax}
\def\prop{\obsolete\prop \propto }
\def\nextequation#1{\obsolete\nextequation \global\equanumber=#1
   \ifnum\the\equanumber>0 \global\advance\equanumber by 1 \fi}
\def\BOXITEM{\afterassigment\B@XITEM\setbox0=}
\def\B@XITEM{\par\hangindent\wd0 \noindent\box0 }
\def\phyzzx{PHY\setbox0=\hbox{Z}\copy0 \kern-0.5\wd0 \box0 X}
%
%
\everyjob{\xdef\today{\monthname\ \number\day, \number\year}}
        
%


\hoffset=0.2truein
\voffset=0.1truein
\hsize=6truein

\def\CALT#1{\hbox to\hsize{\tenpoint \baselineskip=12pt
	\hfil\vtop{\hbox{\strut CALT-68-#1}
	\hbox{\strut DOE RESEARCH AND}
	\hbox{\strut DEVELOPMENT REPORT}}}}

\def\CALTECH{\smallskip
	\address{California Institute of Technology, Pasadena, CA 91125}}
\def\TITLE#1{\vskip 1in \centerline{\fourteenpoint #1}}
\def\AUTHOR#1{\vskip .5in \centerline{#1}}

\def\ABSTRACT#1{\vskip .5in \vfil \centerline{\twelvepoint \bf Abstract}
	#1 \vfil}

\def\sqr#1#2{{\vcenter{\hrule height.#2pt
      \hbox{\vrule width.#2pt height#1pt \kern#1pt
        \vrule width.#2pt}
      \hrule height.#2pt}}}

\def\section#1#2{
\noindent\hbox{\hbox{\bf #1}\hskip 10pt\vtop{\hsize=5in
\baselineskip=12pt \noindent \bf #2 \hfil}\hfil}
\medskip}

\def\underwig#1{	
	\setbox0=\hbox{\rm \strut}
	\hbox to 0pt{$#1$\hss} \lower \ht0 \hbox{\rm \char'176}}

\def\bunderwig#1{	
	\setbox0=\hbox{\rm \strut}
	\hbox to 1.5pt{$#1$\hss} \lower 12.8pt
	 \hbox{\seventeenrm \char'176}\hbox to 2pt{\hfil}}

\def\MEMO#1#2#3#4#5{
\frontpagetrue
\centerline{\tencp INTEROFFICE MEMORANDUM}
\smallskip
\centerline{\bf CALIFORNIA INSTITUTE OF TECHNOLOGY}
\centerline{\tencp Charles C. Lauritsen Laboratory of High Energy Physics} 
\bigskip
\vtop{\tenpoint \hbox to\hsize{\strut \hbox to .75in{\caps to:\hfil}
\hbox to3in{#1\hfil}
\hbox to .75in{\caps date:\hfil}\quad \the\date\hfil}
\hbox to\hsize{\strut \hbox to.75in{\caps from:\hfil}\hbox to 2in{#2\hfil}
\hbox{{\caps extension:}\quad#3\qquad{\caps mail code:\quad}#4}\hfil}
\hbox{\hbox to.75in{\caps subject:\hfil}\vtop{\parindent=0pt
\hsize=3.5in #5\hfil}}
\hbox{\strut\hfil}}}


%
%
%
%
%
%
%



\newread\figureread                                     
\def\g@tfigure#1#2 {\openin\figureread #2.fig           
   \ifeof\figureread \errhelp=\disabledfigures          
     \errmessage{No such file: #2.fig}\let#1=\relax \else
    \read\figureread to\y@p \read\figureread to\y@p     
    \read\figureread to\x@p \read\figureread to\y@m     
    \read\figureread to\x@m \closein\figureread         
    \xdef#1{\hbox{\kern-\x@m truein \vbox{\kern-\y@m truein
      \hbox to \x@p truein{\vbox to \y@p truein{        
        \special{pos,inc=#2.fig}\vss }\hss }}}}\fi }    

\def\section#1{\par\ifnum\the\lastpenalty=30000\else
        \penalty-200\vskip\sectionskip\spacecheck\sectionminspace\fi
        \global\advance\sectionnumber by 1 
        \xdef\sectionlabel{\the\sectionstyle\the\sectionnumber}
        \wlog{\string\section\space\sectionlabel}
        \TableOfContentEntry s\sectionlabel{#1}
        \noindent {\caps\enspace\sectionlabel\quad #1}\par
        \nobreak\vskip\headskip\penalty 30000 }


\input epsf
\ifx\epsffile\undefined\message{(FIGURES WILL BE IGNORED)}
\def\insertfig#1#2{}
\else\message{(FIGURES WILL BE INCLUDED)}
\def\insertfig#1#2{{{\baselineskip=4pt
\midinsert\centerline{\epsfxsize=\hsize\epsffile{#2}}{{\centerline{#1}}}\
\medskip\endinsert}}}

\def\insertfigpage#1#2{{{\baselineskip=4pt
\pageinsert\centerline{\epsfysize=6.5in\epsffile{#1}}
 {{\bigskip{#2}}}\
\smallskip\endinsert}}}

\def\insertcomfigpage#1#2{{{\baselineskip=4pt
\pageinsert\centerline{\epsfysize=6.0in\epsffile{#1}}
 {{\smallskip{#2}}}\
\endinsert}}}

\def\insertxfigpage#1#2{{{\baselineskip=4pt
\pageinsert\centerline{\epsfxsize=6.0in\epsffile{#1}}
 {{\bigskip{#2}}}\
\medskip\endinsert}}}

\def\bcap#1{{\bf Figure {#1}:\/}}

\fi

\tolerance 10000

\def\rhotwiddle{\tilde{\rho}}
\def\Btwiddle{\tilde{B}}
\def\ktwiddle{\tilde{\kappa}}
\def\thtwiddle{\tilde{\theta}}
\def\half{{1\over 2}}

\parindent=20.0pt
\hfuzz=5pt
\tolerance=10000



\CALT{2058}
\TITLE{ Mean Field Theory for Fermion-based $U(2)$ Anyons}
\AUTHOR{Patrick McGraw\footnote{\dag}{E-mail: pmcgraw@theory.caltech.edu}}
\CALTECH

\ABSTRACT{The energy density is computed for a U(2) Chern-Simons theory coupled to a non-relativistic fermion field (a theory of ``non-Abelian anyons'') under the assumptions of uniform charge and matter density.  When the matter field is a spinless fermion, we find that this energy is independent of the two 
Chern-Simons coupling constants and is minimized when the non-Abelian charge density is zero.  This suggests that there is no spontaneous breaking of the $SU(2)$  subgroup of the theory's symmetry, at least in this mean field approximation. For spin-$1/2$ fermions, we find evidence of ground states with a small non-Abelian charge density, which vanishes as the theory of free fermions is approached.}

\section{Introduction}
In this paper, we consider a $2+1$ dimensional model with a non-relativistic matter field $\Psi$ minimally 
coupled to an $SU(2)\times U(1)=U(2)$ ``statistical'' or Chern-Simons gauge field.
The matter will be taken to carry a unit of $U(1)$ charge and a fundamental
representation of $SU(2)$  ``isospin.''  The Lagrangian is given by:
$${\cal L}=i\Psi^{\dagger}(D_0\Psi)-{1\over {2 m}}(D_i\Psi)^{\dagger}
   (D_i\Psi)+{\ktwiddle\over2}\epsilon^{\alpha\beta\gamma}
   (A^a_\alpha \partial_\beta A^a_\gamma-{1\over 3}\epsilon_{abc}
   A^a_\alpha A^b_\beta A^c_\gamma)+{\kappa\over 2}\epsilon^{\alpha
   \beta\gamma}A_\alpha \partial_\beta A_\gamma, \eqn\Lagrangian$$
where $A_\mu$ and $A_\mu^a$ are the $U(1)$ and $SU(2)$ gauge fields, with 
Chern-Simons coupling constants $\kappa$ and $\ktwiddle$ respectively, and 
the covariant derivative is 
$$D_\mu = \partial_\mu+iA_\mu+iA^a_\mu {{\sigma^a} \over 2}.\eqn\deriv$$
(It will sometimes also be convenient to define ${\cal A}_\mu\equiv 
A_\mu + A^a_\mu \sigma^a/2$.)
We will consider cases of either bosonic or fermionic matter fields, so
that $\Psi$ may obey either canonical commutation or anticommutation relations.
The Hamiltonian corresponding to \Lagrangian \/ is given by
$$H=\int d^2{\bf x}{1\over{2 m}}(D_i\Psi)^\dagger(D_i\Psi)\eqn\ham$$
while the gauge fields are subject to constraints which
relate them to the matter fields:
$$ B\equiv -F_{12}=\epsilon_{ij}\partial_iA_j=-{1\over\kappa}\rho,$$
$$ B^a\equiv -F^a_{12}=\epsilon_{ij}(\partial_iA_j^a
   +\half\epsilon^{abc}A_i^bA_j^c)=-{1\over\ktwiddle}\rho^a,\eqn\gauss$$
and
$$ F_{0i}=-{1\over\kappa}\epsilon{ij}J_j,$$
$$ F^a_{0i}=-{1\over\ktwiddle}\epsilon_{ij}J^a_j, \eqn\constraints$$
where the densities and currents are defined by:
$$\rho=\Psi^\dagger\Psi,\/ \rho^a=\Psi^\dagger{\sigma^a\over 2}\Psi,$$
$$J_i={1\over 2 i m}(\Psi^\dagger D_i\Psi-(D_i\Psi)^\dagger \Psi),\/
J_i^a={1\over 2 i m}(\Psi^\dagger {\sigma^a \over 2} D_i\Psi-
    (D_i\Psi)^\dagger  {\sigma^a \over 2} \Psi).\eqn\currents$$
Since there is no Maxwell term, the gauge fields are non-dynamical and are
completely determined (up to gauge transformation) by \gauss \/ and \constraints.

As usual, the theory is gauge invariant only when $\ktwiddle$ is an integer 
multiple of $1/4\pi$.  The Gauss constraints ensure that any electrically charged
particle also carries magnetic flux,  so the particles experience mutual
Aharonov-Bohm interactions resulting in exotic statistics.  In this case,
the braiding of two particles leads not only to an overall phase change of the 
wave function but also a non-Abelian rotation acting on the isospin indices 
of the two particles.  A two-particle total wave function $\psi(x_1,x_2)$ is acted on  by the braiding operator
$$ \exp[i(\theta+{\bf \tilde{\Theta}})] \equiv
    \exp[{-i\over{2\kappa}}+{i{\bf\sigma_{(1)}\cdot\sigma_{(2)}}\over\ktwiddle}],
    \eqn\monodromy$$
where ${\bf \tilde{\Theta}}$ can be thought of as a matrix-valued $SU(2)$ ``phase.''  The 
eigenstates of this two-particle braiding operator are states of definite total
isospin, i.e., pure iso-singlet or triplet states. 

 \REF\VC{A. Cappelli and P. Valtancoli, Nuc. Phys. {\bf B 453}, 727 (1995).}
This model of ``non-Abelian anyons'' was studied in reference [\VC] using a mean field approximation.  In this mean field technique
one first quantizes the matter field in the presence of a classical background magnetic field.  One searches for a self-consistent ground state having a uniform expectation value of the matter and isospin densities consistent with the Gauss constraints.   Then, using the densities and currents as fundamental variables, one can study the effect of fluctuations about this approximate ground state, using the Bogoliubov approximation in which the commutators of fluctuation operators are replaced by their expectation values in the zeroth-order ground state.\Ref\Trugenberger{C.A.
Trugenberger,  Phys. Rev. {\bf D 45}, 3807 (1992).}

In reference [\VC ], it was shown that with bosonic matter the mean field ground state energy behaves differently depending on the values of the two Chern-Simons couplings.
In particular, there was a phase in which the energy was minimized
in the mean field approximation by the generation of a non-zero isospin density, and thus the $SU(2)$ symmetry was spontaneously broken.  This paper will investigate the consequences of
coupling fermions instead of bosons to the $U(2)$  Chern-Simons field.  We 
will show that for spinless (or polarized) fermions, the resulting mean field theory differs from the bosonic case
in two ways:  (1)  The form of the mean field energy density as a function of 
matter and charge densities is independent of the Chern-Simons coupling constants and (2)  For a given particle density, the lowest energy always 
occurs when the $SU(2)$ charge density is zero.  Thus the mean field picture
does not show any breaking of the $SU(2)$ symmetry.  In the case of spin-1/2
fermions, however, there is a hint of spontaneous symmetry breaking which vanishes smoothly as the theory approaches that of pure fermions.

\section{Mean field approximation for fermion-based U(2) anyons}

For a theory where the matter field $\Psi$ is fermionic, we now search, much as in reference [\VC ], for a
self-consistent ground state $\ket{\Omega}$ with uniform matter and isospin densities:
$$\VEV{\rho}=\rho_0,\/\/ \VEV{J_i}=0,$$
$$ \VEV{\rho^a}=\rho_0^a,\/\/ \VEV{J_i^a}=0.\eqn\densities$$
According to the Gauss constraints, these nonzero densities will imply uniform magnetic fields.  We will consider matter in the background of these fields, and then demand that the number of particles per unit area be consistent with the
Gauss constraints. 
We can take the isospin density to be along the $\sigma_3$ direction, 
$\VEV{\rho^a}=\delta^a_3 \rhotwiddle_0.$  Then we have $U(1)$ and $SU(2)$
magnetic fields given by 
$$\VEV{B}=B_0=-{1\over\kappa}\rho_0,\/\/
  \VEV{B^a}=\delta^a_3 \Btwiddle_0=-{1\over\ktwiddle}\delta^a_3\rhotwiddle_0.
  \eqn\bfields$$
In the symmetric (or isotropic) gauge\Ref\isotropic{See, for example, E. Fradkin,  Field Theories of Condensed Matter Systems
(Addison-Wesley, New York, 1991).} ${\cal A}_i=(B/2)\epsilon_{ij}x^j$, the gauge field can be written as
$${\cal A}_i=\epsilon_{ij}{x^j\over 2}({\rho_0\over\kappa}+
      {\rhotwiddle_0\over{2 \ktwiddle}}\sigma_3).\eqn\gaugefield$$

The single-particle orbitals are split into two sets according to the
eigenvalue of $\sigma_3$.  Particles in states of up or down isospin feel
different effective magnetic fields,  $B_+=(B_0+\Btwiddle_0/2)$ and 
$B_-=(B_0-\Btwiddle_0/2)$,  respectively.  The single-particle energy levels thus fall into two sets of Landau levels.  The kth energy levels of the isospin up and down systems have energies given by 
$$ \epsilon_k^{\pm}={B_\pm\over m}(k+\half),\eqn\levels$$
and have degeneracy per unit area $N_{\pm}/A=|B_{\pm}|/2\pi.$

Uniformity of the density requires that all orbitals of any given Landau level be filled with the same number of particles.\Ref\uniform{Some subtleties are being hidden within the phrase ``all orbitals of any given Landau level.''
A droplet of finite area, uniform everywhere inside and dropping sharply at the edges, is generated
by filling a finite number of degenerate orbitals, and a consideration of the 
droplet's edges is necessary when fluctuations about the mean-field state are
taken into account.  See  G.V.~Dunne, Int. J. Mod. Phys {\bf B 8}, 1625 (1994).}  In the case of spinless (or spin-polarized) fermions,
this means either 0 or 1 per orbital.  For spin-$1/2$ fermions, the possible occupancies are 0,1, and 2.  Let us first consider spinless fermions.  The total energy per unit area for a state with
the lowest n levels filled is given by:
$${E \over A}={B\over{2 \pi}}\sum_{k=0}^{n-1}\epsilon_k={B^2n^2\over{4\pi m}}.\eqn\halfE$$
(The factor $B/2\pi$ in front of the sum represents the degeneracy.)  In our
system,  if the lowest $n_+$ and $n_-$ of the isospin up and down levels, respectively, are uniformly filled with one particle per orbital, then the combined energy is given by
$${E\over A}={n_+^2\over{4 \pi m}}(B_0+{\Btwiddle_0\over 2})^2
            + {n_-^2\over{4 \pi m}}(B_0-{\Btwiddle_0\over 2})^2 .\eqn\fullE$$
The density of isospin-up particles is given by the number of filled Landau
levels times the degeneracy, $\rho_+ = n_+N_+/A$, and similarly for the down particles it is $\rho_- = n_-N_-/A$.  The matter density $\rho$ is the sum of the up and down densities, while the isospin density for isospin-1/2 particles
is given by half the difference\refmark\VC:
$$\rho_0=\rho_+ +\rho_- = {n_+N_+\over A}+{n_-N_-\over A}=
          n_+{{|B_0+\Btwiddle_0/2|}\over{2 \pi}}+
          n_-{{|B_0-\Btwiddle_0/2|}\over{2 \pi}},$$
$$ \rhotwiddle_0=\half(\rho_+ - \rho_-)
     = \half( n_+{{|B_0+\Btwiddle_0/2|}\over{2 \pi}}-
                      n_-{{|B_0-\Btwiddle_0/2|}\over{2 \pi}}).\eqn\densities$$

These equations may be rewritten in the form 
$${n_+|B_0+\Btwiddle_0/2|\over{2\pi}}={\rho_0\over 2}+\rhotwiddle_0,\/\/
  {n_-|B_0-\Btwiddle_0/2|\over{2\pi}}={\rho_0\over 2}-\rhotwiddle_0.
  \eqn\altdensities$$
The expression \fullE\/ for the energy then becomes:
$${E\over A}={\pi\over{m}}[({\rho_0\over 2}+\rhotwiddle_0)^2+
                        ({\rho_0\over 2}-\rhotwiddle_0)^2]=
             {2\pi\over m}({\rho_0^2\over 4}+\rhotwiddle_0^2).\eqn\finalE$$
For comparison, the result in the bosonic case of [\VC\/] was found to be:
$${E\over A}={1\over 2 m}\left[ |{\rho_0\over \kappa}+{\rhotwiddle_0\over 2\ktwiddle}|
               ({\rho_0\over 2}+\rhotwiddle_0)+
              |{\rho_0\over \kappa}-{\rhotwiddle_0\over 2\ktwiddle}|
               ({\rho_0\over 2}-\rhotwiddle_0)\right].\eqn\bosonicE$$

The latter expression is different because in the bosonic case, only the lowest  of each of the two sets of Landau levels is occupied in the ground state, whereas in the 
fermionic system, the exclusion principle requires that higher levels be occupied.
The two expressions \finalE \/ and \bosonicE\/ are plotted in figure \FIG\spinZplot~\spinZplot \/ for 
representative values of the coupling constants.

\insertcomfigpage{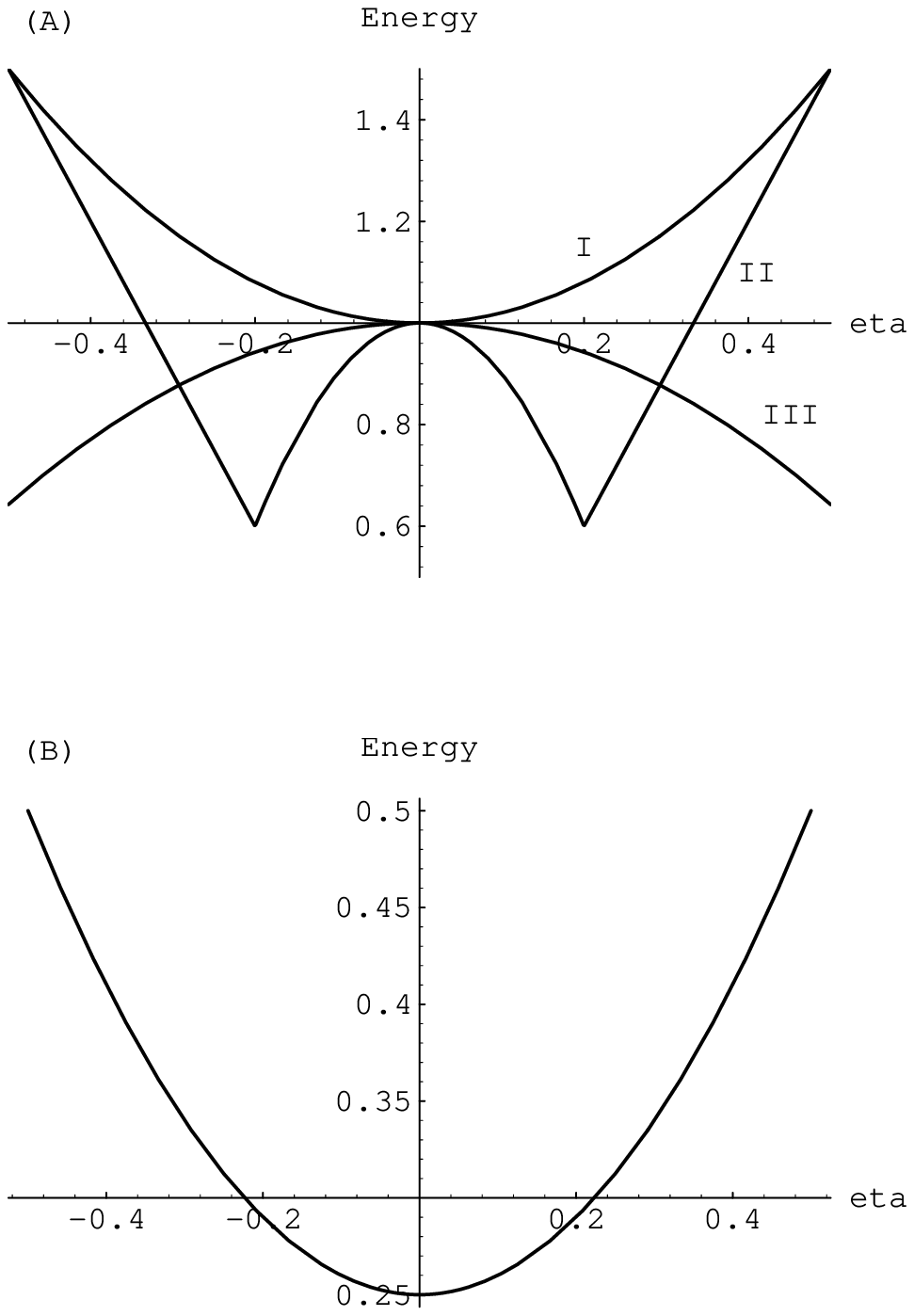}{\bcap{\spinZplot}  \/ Upper plot (A) (adapted from reference [ \VC ]) : energy of bosonic mean-field theory plotted against the ratio $\rhotwiddle_0 / \rho_0 $  for three values
of the coupling constants.  Curve I is for $\ktwiddle/\kappa=0.5$,  and is
typical of the region $\kappa \ktwiddle >0$. Curve II, with 
$\ktwiddle/\kappa=-0.1$,  is typical for $\kappa \ktwiddle <0 $ and $|\ktwiddle| / |\kappa| <.25$.  Curve III, with $\ktwiddle/\kappa=-0.7$, is representative of
$\kappa \ktwiddle <0$, $|\ktwiddle| / |\kappa| >0.25$.  On the x-axis is the ratio 
$\rhotwiddle_0/\rho_0$, on the y-axis is energy in units of $1/2m\kappa$. In case
I, The minimum occurs at $\rho_0=0$.  In case III there is a self-consistent
mean-field ground state with maximal isospin density, $\rhotwiddle_0=\rho_0/2$
In case II, the mean-field approximation is not self-consistent:  The minimum 
energy occurs in a limit where one of the effective magnetic fields $B_+$ or
$B_-$ goes to zero and the Landau level picture breaks down.  The lower graph
(B) shows the much simpler form of the energy expression for spinless fermions: it
is independent of $\kappa$ and $\ktwiddle$, and is always a minimum when 
$\rhotwiddle_0=0$. The energy scale for the lower plot is $\pi/2 m$.}

We note the following features of the fermionic result which differ from those of the bosonic result:
(1)  The expression is completely independent of the Chern-Simons coupling 
constants $\kappa$ and $\ktwiddle$, and was derived without any reference to
the Gauss constraints $-\kappa B_0=\rho_0$ and $-\ktwiddle\Btwiddle_0=
\rhotwiddle_0$.  (2)  For a given matter density $\rho_0$, the energy is always
minimized by $\rhotwiddle_0=0$. Thus it appears that the assumptions
\densities\/ for the ground state are not self-consistent unless 
$\rhotwiddle_0=0$, and there is, in this approximation, no spontaneous breaking of the $SU(2)$ symmetry.  

The first of
the above observations is not entirely surprising in view of the results for the Abelian model (which should correspond to the limit
$1/|\ktwiddle|\ll 1/|\kappa|$).  When $\rhotwiddle_0=0$, \finalE\/ reduces to
the result for fermion-based Abelian anyons.  This result is similarly 
independent of the Chern-Simons coupling, and is equal to the energy of a degenerate fermi gas in $2+1$ dimensions.  Corrections to this fermi 
energy are found only when one includes the effects of fluctuations to quadratic order.  The second property, the absence of a spontaneous non-Abelian charge
density, is less obvious.  In the boson-based case, it was found that the 
phase with $\rhotwiddle_0=\pm\rho_0/2$ was stable in the regime  
$1/|\ktwiddle|< 4/|\kappa|$ with $\ktwiddle$ and $\kappa$ having opposite signs.
Fermions can in principle be generated from bosons by setting the Abelian
Chern-Simons coupling to $\kappa=-1/2\pi$, resulting in a statistical angle of $\pi$.  If we were to continue naively from the behavior near the bosonic point ($\kappa \to \infty$) to $\kappa=-1/2\pi$,
we might expect symmetry breaking at the fermionic end when
$|\ktwiddle|>{1\over{8\pi}}$.  It might be argued, in view of the spin-statistics connection, that it is not natural to expect a theory of spinless
bosons to be connected continuously to one of {\it spinless} fermions.  The
properties of the pure Abelian theory obtained with the prescription of refs. [\VC ] and [\Trugenberger ] interpolate smoothly  between the bosonic theory and a theory of unpolarized
spin-1/2 fermions.\Ref\otherspin{This is not the only possible prescription:
theories of spinless anyons can also be constructed.  For example, see 
K.H. Cho, C. Rim, D.S. Soh, Phys. Lett. {\bf A 164}, 65 (1992); Y.H. Chen, F. Wilczek, E. Witten, B.J. Halperin, Int L. Mod. Phys {\bf B 3}, 1001 (1989);
A.L. Fetter, C.B. Hanna, R.B. Laughlin, Phys. Rev. {\bf B 39}, 9679 (1989).}  Therefore, in the next section, we will consider a spin-1/2
matter field.

Qualitatively, the comparatively greater susceptibility of the bosonic model to the development of a spontaneous asymmetry between the isospin-up and down Landau levels may be explained in terms of the particles' exclusion properties.  An isospin asymmetry generally raises the energies of one set of Landau levels while lowering those of the other.  Since any number of bosons may occupy the lowest Landau level,  it can become energetically favorable to lower the energy of, say, the isospin-up Landau level and place all of the particles into this lowered level, while leaving the raised level unoccupied.  The fermionic system, on the other hand, can be regarded to a first approximation as two fermi fluids, one filling the isospin-up levels and one filling the down levels.  The development of an isospin asymmetry requires that more particles be added to one
of the two fluids.  Even if energy of the kth Landau level is lowered by the 
asymmetry, the levels must be filled up to a higher value of k, which offsets the energetic advantage.

\section{Spin-1/2 fermions}
We now ask whether the result of the previous section is changed if we consider
spin-1/2 fermions instead of spinless ones.  For spin-1/2 fermions, there are
two states per orbital,  and so it is possible to fill a Landau level with either 1 or 2 particles per orbital.   (We are supposing that the statistical gauge field does not couple to the spin, so that this double occupation is the only effect.)  Let $n=2p+\sigma,\/ \sigma=0,1$, and consider a state in which the lowest
$n/2$ of a set of Landau levels are filled.   If $n$ is odd ($\sigma=1$), we mean by this that the lowest $p$ levels are doubly filled, and the $p+1$ level is filled with one particle per orbital.  The total
energy per unit area of such a configuration is \refmark\VC
$${E\over A}={B\over 2\pi}[2\sum_{k=0}^{p-1}{B\over m}(k+\half)+
      \sigma{B\over m}(p+\half )]={B^2\over 8 \pi m}(n^2+\sigma).\eqn\spinhalf$$

For our system with two sets of Landau levels, the total energy becomes
$${E\over A}={(n^2_++\sigma_+)\over 8 \pi m}(B_0+{\Btwiddle_0\over 2})^2+
    {(n^2_-+\sigma_-)\over 8 \pi m}(B_0-{\Btwiddle_0\over 2})^2.\eqn\fullspinE$$
As before, the $+$ and $-$ subscripts refer to the isospin states.
Whereas the corresponding equation \fullE\/ for the spinless case only involved
the products  $n_\pm|B_0\pm {\Btwiddle_0\over 2}|$ and thus could be expressed
in terms of the densities without reference to the Gauss constraints, that is 
not the case here.  Using the Gauss constraints, we write 
$$B_0\pm{\Btwiddle_0\over 2}=
    -{\rho_0\over \kappa}\mp{\rhotwiddle_0\over{2\ktwiddle}},$$
and thus
$${E\over A}={\pi\over m}[({\rho_0^2\over 4}+\rhotwiddle_0^2)
             +{\sigma_+\over 8 \pi^2}
              ({\rho_0\over\kappa}+{\rhotwiddle_0\over 2 \ktwiddle})^2
             +{\sigma_-\over 8 \pi^2}
             ({\rho_0\over\kappa}-{\rhotwiddle_0\over 2 \ktwiddle})^2].
\eqn\a$$
Since the second and third terms are non-negative, we see that, at the mean-field level, the lowest energy state for a given 
$\rho_0$ is still one with $\sigma_+=\sigma_-=0$ 
and $\rhotwiddle_0=0$.  However, if $\sigma_+$ or $\sigma_-$ is restricted to
be 1, then a minimum of the energy does in fact appear at
$$\rhotwiddle_0=\pm{2\ktwiddle\over\kappa}
    ({1\over{32\pi^2\ktwiddle^2+1}})\rho_0
     \approx{\pm \rho_0\over{16\pi^2\kappa\ktwiddle}}.\eqn\newmin$$
(See figure \FIG\spinhalfplot\/ \spinhalfplot .)
This expectation value of $\rhotwiddle$ vanishes in the limit of free fermions
($1/\kappa\rightarrow 0, 1/\ktwiddle\rightarrow 0$).
The corresponding energy, in the limit where both $1/\kappa$ and $1/\ktwiddle$ 
are small, is given by
$${E\over A}={\pi\over m }[{\rho_0^2\over 4} + {\rho_0^2\over{8\pi^2\kappa^2}}+
    {\cal O}({1\over\kappa^4})].\eqn\newminE$$
  It is worth noting that in
the Abelian case, states with the top level half-filled are also energetically
unfavorable at the mean-field level, but that fluctuations introduce $\sigma$-
dependent corrections of the same approximate size (${\cal O}(1/\kappa^2)$)
and {\it opposite sign}.  Thus it is conceivable that in our model, the 
$\sigma_\pm=1$ ground states might be stabilized by quadratic corrections.  Also, in the Abelian model,  the half-filled ground states are the only consistent ones at odd values of the coupling constant.

\insertfigpage{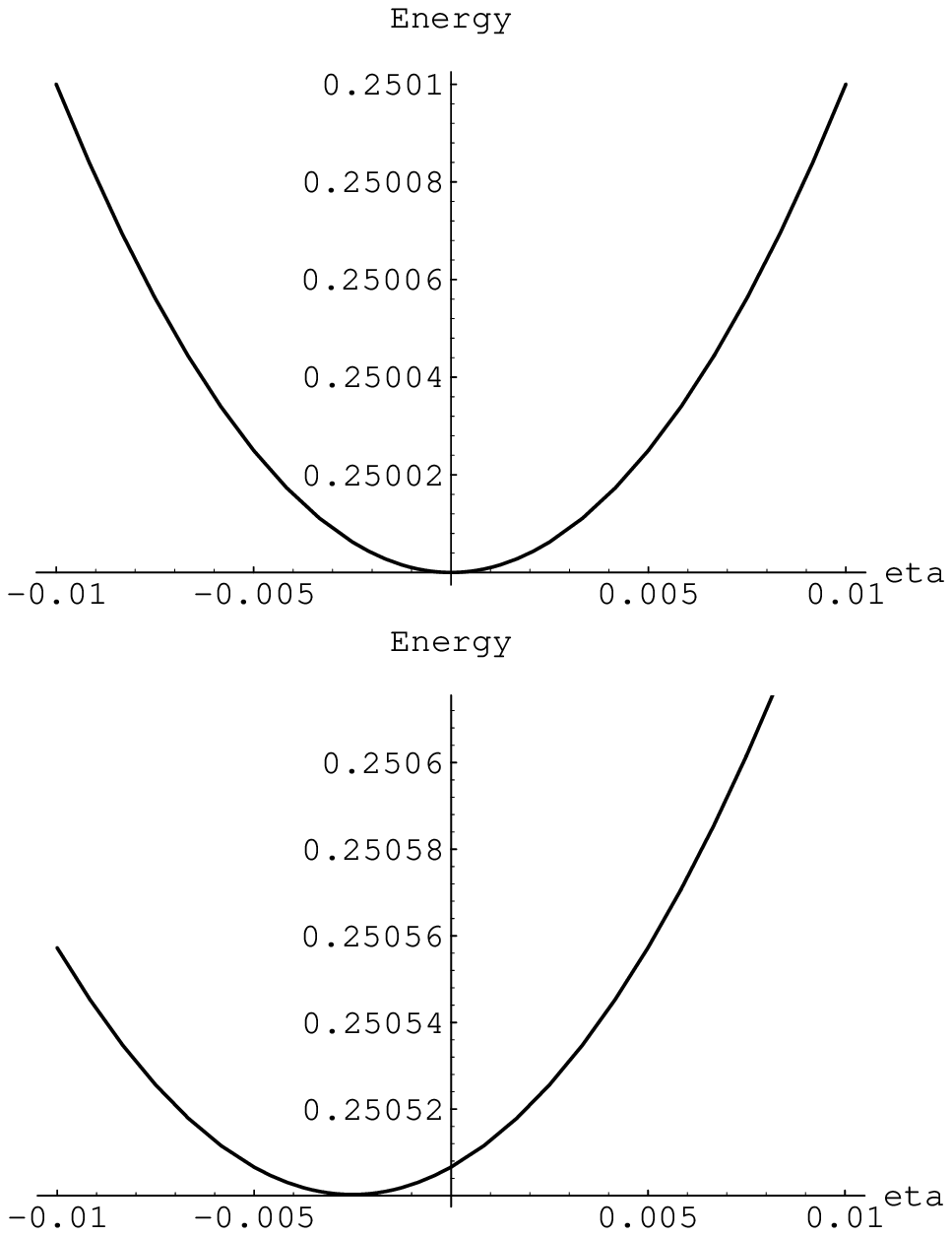}{\bcap{\spinhalfplot \/ A}  \/ Energy of fermion-
based anyons for $\sigma_+=\sigma_-=0$ (top) and for
$\sigma_+=1, \/ \sigma_-=0$  (bottom).}

\insertfigpage{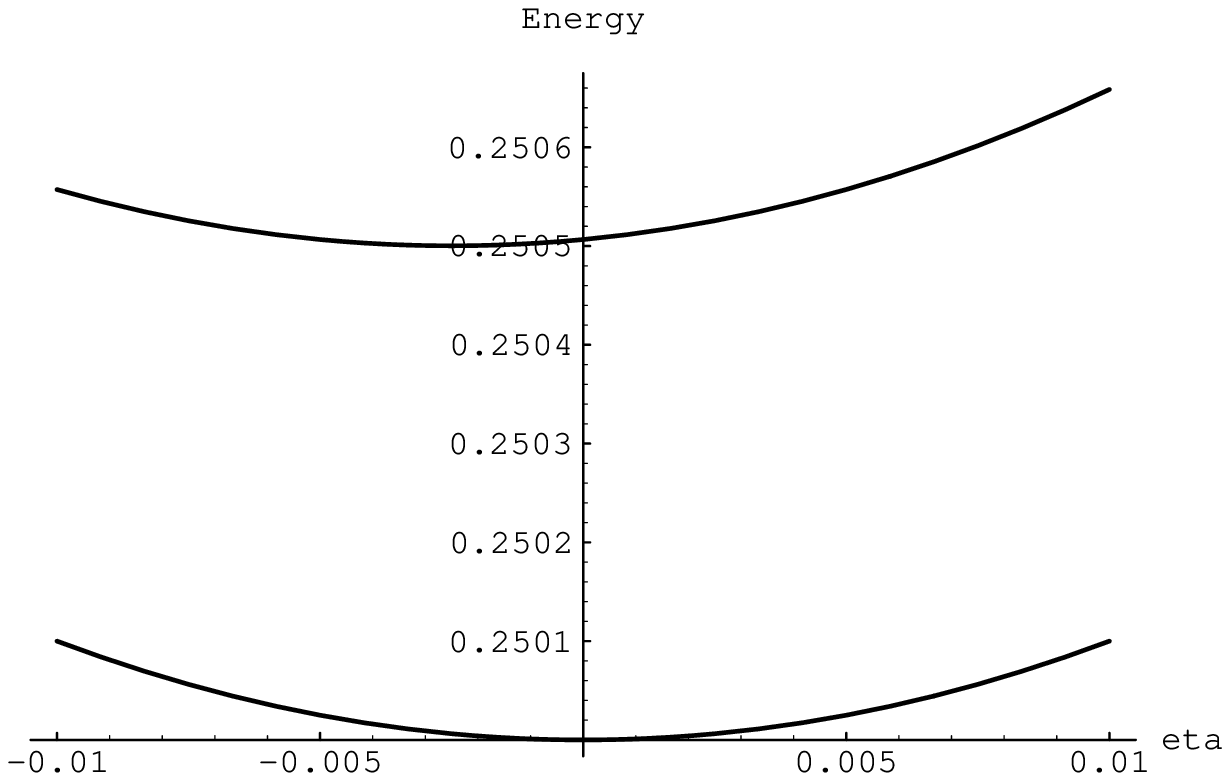}{\bcap{\spinhalfplot \/ B} \/ The two curves of 
figure \spinhalfplot\/ A, shown on the same axes.}

\section{Consistency of mean-field theories}
The assumption of a uniform matter density in a uniform magnetic field requires
that all orbitals within a Landau level be filled equally.  This means that the only degree of 
freedom for the ground-state distribution of particles in a set of Landau levels is the filling factor, which must be an integer.  In the case of a bosonic ground state, this integer represents the occupation number of all orbitals
in the lowest Landau level, while in a fermionic ground state it represents
the number of levels which are filled. Combined with the Gauss constraints,
this condition will in general pick out a discrete set of values of the Chern-Simons coupling constants at which the mean-field ground state is well-defined.  In the case of Abelian anyons (either boson- or fermion-based)
there is only one set of Landau levels which may be filled with an integer factor.  The matter density is given by the filling factor times the Landau
level degeneracy per unit area.  The degeneracy is related to the magnetic
field, which is in turn related to the density by the Gauss constraint:
$$\rho_0={{n |B_0|}\over{2 \pi}} =
    {n\over2\pi}|{\rho_0\over \kappa}|.\eqn\abseries$$
$\rho_0$ may be divided out from both sides, leading to the familiar series of
allowed values of the coupling constant: $\kappa=n/2\pi$, corresponding to the
series of statistical angles $2\pi/n$ and $1-2\pi/n$  based on bosons and fermions, respectively.  

Similar constraints occur in the non-Abelian model we are considering, but the consistency conditions are more complicated because two sets of Landau levels are involved.  This issue was not addressed in reference [\VC ], so we consider 
here both the bosonic and fermionic systems.  The basic equations are those 
of \densities , which relate the isospin and matter densities to the filling factors and magnetic fields.  When the Gauss constraints \gauss\/ are applied to substitute for the 
magnetic fields in \densities , these equations become:
$$\rho_0={n_+\over{2\pi}}|{\rho_0\over\kappa}+{\rhotwiddle_0\over{2\ktwiddle}}|
  +{n_-\over{2\pi}}|{\rho_0\over\kappa}-{\rhotwiddle_0\over{2\ktwiddle}}|,$$

$$\rhotwiddle_0=
   {n_+\over{2\pi}}|{\rho_0\over\kappa}+{\rhotwiddle_0\over{2\ktwiddle}}|
  -{n_-\over{2\pi}}|{\rho_0\over\kappa}-{\rhotwiddle_0\over{2\ktwiddle}}|.
   \eqn\basiceqns$$
These basic equations have the same form for fermi and bose-based systems; only the interpretation of the filling factors $n_+$ and $n_-$ is different.

  Now 
consider a mean-field state with 
$$\rhotwiddle_0={\eta\over 2}\rho_0,$$
where $-1<\eta<1$.
In such a state, equations \basiceqns \/ become:
$$ 1 = {n_+\over{2\pi}}|{1\over\kappa}+{\eta\over{4\ktwiddle}}|+
       {n_-\over{2\pi}}|{1\over\kappa}-{\eta\over{4\ktwiddle}}|  $$
$$ \eta =  {n_+\over{2\pi}}|{1\over\kappa}+{\eta\over{4\ktwiddle}}|-
       {n_-\over{2\pi}}|{1\over\kappa}-{\eta\over{4\ktwiddle}}|.\eqn\basica$$
Because of the quantization of $\ktwiddle$ we may write $\ktwiddle=m/4\pi$,
where $m$ is an integer.
With this substitution, the equations \basica\/ may be combined into a relation between $n_+$ and $n_-$, and another relation involving only $n_+$:
$$(1-\eta){n_+\over{2\pi}}|{1\over\kappa}+{\eta\pi\over m}| =
  (1+\eta){n_-\over{2\pi}}|{1\over\kappa}-{\eta\pi\over m}|,\eqn\comba$$
$$\pi{{1+\eta}\over n_+} = |{1\over\kappa}+{\eta\pi\over m}|.\eqn\combb$$
In the case $\eta = 0$, these reduce to $\kappa={n_+\over\pi}={n_-\over\pi}$,  reproducing the
familiar set of Abelian mean-field theories.
The bosonic ground state of maximal isospin alignment described in ref. [\VC]
corresponds to $\eta=1$.  In this case, we have $n_-=0$ and
$${1\over\kappa}={2\pi\over n_+}-{\pi\over m},\eqn\bosetower$$
and thus find that $\kappa$ takes values which are rational, but not necessarily integer, multiples of ${1/\pi}$.  $\kappa$ approaches integer values only in 
the limit $m\gg n_+$, or $\ktwiddle\gg \kappa$.

We now consider the mean-field energy minimum \newmin \/ of the spin-1/2 fermion-based theory, which corresponds to 
$$\eta = {-4\ktwiddle\over\kappa}\left({1\over{32\pi^2\ktwiddle^2+1}}\right)
       = {-m\over{\pi\kappa}}\left({1\over{2 m^2+1}}\right).\eqn\smalleta$$  The conditions on the 
coupling constants for a consistent mean-field theory at this value of $\eta$
turn out to be more complicated.  Noting the useful expressions 
$$1\pm\eta={{\pi\kappa(2m^2+1)\mp m}\over{\pi\kappa(2m^2+1)}}\eqn\usefula$$
and
$$|{1\over\kappa}\pm{\eta\over{4\ktwiddle}}|=
  {1\over{|\kappa|}}\left({{2m^2+1\mp1}\over{2m^2+1}}\right),\eqn\usefulb$$
we find that equation \comba \/ (which relates $n_-$ and $n_+$) takes the form
$$n_+[\pi\kappa(2m^2+1)+m]m^2=n_-[\pi\kappa(2m^2+1)-m](m^2+1).\eqn\combahere$$
Note that, as usual, $\pi\kappa$  must be rational; we may write $\pi\kappa=p/q$,  where $p$ and $q$ are relatively prime integers.  By assumption, $n_-$ is even while $n_+$ is odd.  When equation \combahere \/ is multiplied by $q$,  the RHS
is even, so that\/ $n_+[p(2m^2+1)+qm]m^2$ \/ must likewise be even.  This can only be satisfied if either $m$ is even, or $p$,$q$, and $m$ are all odd.

If $\kappa>0$, the other consistency equation \combb \/ becomes, after substitution of the expressions \usefula \/ and \usefulb:
$$\pi\kappa(2m^2+1)-m=2m^2n_-,$$
or
$$\pi\kappa={{2m^2n_+ +m}\over{2m^2+1}}.\eqn\combbhere$$
Substitution of the above expression for $\pi\kappa$ in \combahere \/ yields:
$$m(mn_+ +1)=n_- (m^2+1).\eqn\combaherea$$
Letting $n_+=n_- +D$, with $D$ an odd integer, we find 
$$D(m^2+1)+m=n_+.\eqn\a$$
By substituting this into the expression \combbhere \/ for $\pi\kappa$, we finally
obtain a relation between the two coupling constants:
$$\pi\kappa = {{2m^2((Dm^2+1)+m)}\over{2m^2+1}}$$
$$\approx Dm^2 =16\pi^2D\ktwiddle^2,\eqn\weirdrel$$
the latter expression being valid in the limit of small inverse coupling constants ${1\over\kappa},{1\over\ktwiddle}\to 0$ \/ (near the point of free fermions). 

\section{Discussion and Conclusions}
The relation \weirdrel \/ describing the coupling constants at which ground states with broken $SU(2)$ occur is a rather peculiar one. It is nonlinear, 
and does not appear to connect continuously to the theory's behavior near the 
bosonic point.  Translated into the variables $\theta\equiv 1/\pi\kappa$ and 
$\thtwiddle\equiv 1/\pi\ktwiddle$,  which represent the sizes of the Abelian and
non-Abelian statistical phases,  the relation becomes (in the 
$\theta,\thtwiddle\to 0$ limit):
$$\theta={\thtwiddle^2\over{16 D}}.\eqn\altwierdrel$$ 

In figure \FIG\phasediag \/ \phasediag, the behavior of the mean field theory is plotted 
in the $\theta-\thtwiddle$ plane between $\theta=0$ and $\theta=\pi$. The 
points $(0,0)$ and $(\pi,0)$ mark free bosons and free fermions, respectively,
and it is near these two points that one expects the mean-field approximation to
be useful.  The odd-$n$ fermion-based ground states appear between the line
$\theta=\pi$ and a parabola,  and are clustered near the $\theta=\pi$ axis.
There is no obvious way to continue this behavior to that at the bosonic point.

\insertxfigpage{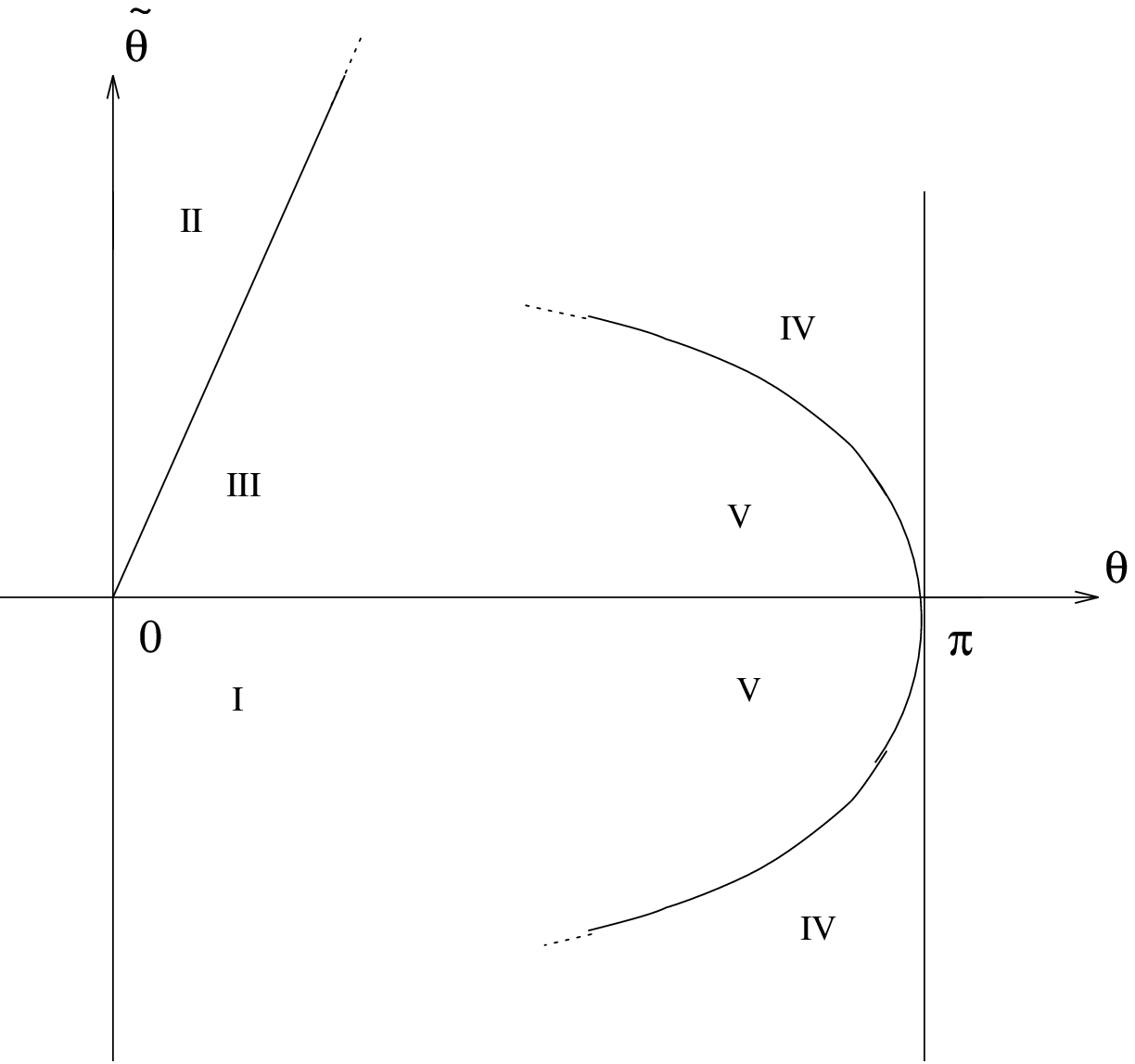}{\bcap\phasediag \/ The $\theta - \thtwiddle$ plane.   The theory's behavior near the points $(0,0)$  (pure bosons) and 
$(0,\pi)$  (pure fermions)  delineates several regions.  In region I, there is no
spontaneous breaking of $SU(2)$:  the energy is minimized by $\rhotwiddle = 0$. 
In region III, there is a consistent mean-field ground state with a maximal
expectation value of $\VEV{\rhotwiddle} = \rho_0/2$.  In region II,  mean-field theory fails:  The ground state energy appears to have a minimum at some 
$\VEV{\rhotwiddle}  < \rho_0/2$, but no self-consistent state can be constructed at that minimum.  These are the three regions of parameter space corresponding to the curves I, II, and III  in figure \spinZplot \/.  Near the fermion point $(0,\pi)$ (regions IV and V), there are mean-field ground 
states with $\VEV{\rhotwiddle} = 0$.  In region IV, there also exist states differing from the others only at quadratic order, and having a small expectation
value of $\rhotwiddle$.  These are the odd filling factor states we have been discussing.  It is not clear how these regions might connect to one another in
areas far from the fermionic and bosonic points.}

One might draw one of two conclusions:  Either there are phase transitions on 
the $\theta - \thtwiddle$ plot other than the ones shown (i.e., between 
$\theta = 0$ and $\theta=\pi$), or mean-field theory 
alone is not sufficient to understand the symmetry-breaking behavior of this theory near the free-fermion point.  The fact that the expectation value 
\newmin \/ of the isospin density is second order in inverse coupling constants
lends credence to the suspicion that the second conclusion is true: mean-field results typically receive corrections at quadratic order in inverse coupling constants due to fluctuations.  The results in this paper, however, suggest that: (1) if there is a spontaneous isospin density near the fermionic point, it vanishes smoothly as that point is approached, and (2) it is worthwhile to 
study the question using other methods.  It has been shown, for example, that 
the pure $SU(2)$ theory (to which our model reduces in the limit 
$\kappa/\ktwiddle\to 0$) is susceptible to the formation of Cooper pairs in an 
isosinglet state.\Ref\kapustin{A. Kapustin, unpublished calculation.}  The condensation of isosinglet Cooper pairs would naturally form a ground state with zero 
isospin density.  But it is possible that when the Abelian coupling is also
included, isotriplet Cooper pairs might form in some regions of parameter space.

\ack

I thank Anton Kapustin for a helpful conversation. I also thank John Preskill,
Hoi-Kwong Lo and John Schwarz for useful discussions and comments on the manuscript.  This work was supported in part by U.S. Department of Energy 
Grant no. DE-FG03-92-ER40701.
  
\par\penalty-400\vskip\chapterskip\spacecheck\referenceminspace
   \ifreferenceopen \Closeout\referencewrite \referenceopenfalse \fi
   \line{\fourteenrm\hfil REFERENCES\hfil}\vskip\headskip
   \input referenc.txa
   
\end